%% file: main.tex
\documentclass[letterpaper,twocolumn,10pt]{article}
\usepackage{usenix2019_v3}

\usepackage[numbers]{natbib}

\usepackage{amsmath}
\usepackage{amssymb}

\usepackage{graphicx}
\usepackage{float}
\usepackage{subfig}

\usepackage{multirow}
\usepackage{booktabs}
\usepackage{xspace}

\usepackage{hyperref}
\usepackage{url}

\usepackage{xcolor}
\usepackage{soul}

\usepackage[linesnumbered,ruled]{algorithm2e}

\usepackage{pifont}

\usepackage[switch]{lineno}

\usepackage{thmbox}

\usepackage[font=bf]{caption}
\usepackage{hyperref}

\newcommand{\ie}{{\it{i.e.}}\xspace}
\newcommand{\eg}{{\it{e.g.}}\xspace}
\newcommand{\vs}{{\it{v.s.}}\xspace}
\newcommand{\etal}{{\it{et al.}}\xspace}
\newcommand{\fig}[1]{Figure~\ref{fig:#1}}
\newcommand{\tab}[1]{Table~\ref{tab:#1}}
\newcommand{\eq}[1]{Equation~\ref{eq:#1}}

\def\name{\textit{Chrion}\xspace}
\begin{document}

\title{\name: Optimizing Recurrent Neural Network Inference by Collaboratively Utilizing CPUs and GPUs}

\author{
{\rm Zinuo Cai}\\
Shanghai Jiao Tong University
\and
{\rm Hao Wang}\\
Louisiana State University
\and
{\rm Tao Song}\\
Shanghai Jiao Tong University
\and
{\rm Yang Hua}\\
Queen's University Belfast
\and
{\rm Ruhui Ma}\\
Shanghai Jiao Tong University
\and
{\rm Haibing Guan}\\
Shanghai Jiao Tong University
}

\date{}
\maketitle

\thispagestyle{empty}

\begin{abstract}

Deploying deep learning models in cloud clusters provides efficient and prompt inference services to accommodate the widespread application of deep learning.
These clusters are usually equipped with host CPUs and accelerators with distinct responsibilities to handle serving requests, \ie~general-purpose CPUs for input preprocessing and domain-specific GPUs for forward computation.
Recurrent neural networks play an essential role in handling temporal inputs and display distinctive computation characteristics because of their high inter-operator parallelism.
Hence, we propose \name to optimize recurrent neural network inference by collaboratively utilizing CPUs and GPUs.
We formulate the model deployment in the CPU-GPU cluster as an NP-hard scheduling problem of directed acyclic graphs on heterogeneous devices.
Given an input model in the ONNX format and user-defined SLO requirement, \name firstly preprocesses the model by model parsing and profiling, and then partitions the graph to select execution devices for each operator.
When an online request arrives, \name performs forward computation according to the graph partition by executing the operators on the CPU and GPU in parallel.
Our experimental results show that the execution time can be reduced by 19.4\% at most in the latency-optimal pattern and GPU memory footprint by 67.5\% in the memory-optimal pattern compared with the execution on the GPU.

\end{abstract}

\input{contents/01_introduction.tex}
\input{contents/02_background.tex}
\input{contents/03_formulation.tex}
\input{contents/04_design.tex}

\input{contents/impl.tex}
\input{contents/05_evaluation.tex}

\input{contents/06_related.tex}
\input{contents/07_conclusion.tex}

\bibliographystyle{plain}
\bibliography{references}

\end{document}

%% file: contents/01_introduction.tex
\section{Introduction}

Prosperous development of deep learning (DL) has brought more and more DL models deployed in cloud computing clusters and providing inference services to users.
For example, Microsoft's Deep-Learning-Inference-Service (DLIS)~\cite{soifer2019deep} serves thousands of machine learning models worldwide and processes three million inference calls per second.
Facebook~\cite{hazelwood2018applied} handles trillions of serving requests to provide user-interactive services, like recommendation and advertising.
The enormous number of concurrent requests puts forward demanding requirements for commercial companies to build the infrastructure for the serving system.
Numerous researchers and engineers have proposed novel techniques to build high-performance deep learning serving systems~\cite{gunasekaran2022cocktail, gujarati2020serving, zhang2019mark, ali2020batch}.

Most current research on serving systems focuses on designing general algorithms to improve the quality of inference services and meet the users' Service Level Objective (SLO).
Batch processing~\cite{ali2020batch,yang2022infless,gujarati2020serving} is commonly adopted when handling requests to improve the system's throughput and reduce the average execution time.
Model compression~\cite{deng2020model, bhardwaj2019memory, zhang2018efficient} is proposed to minimize model size and accelerate inference speed. 
Besides, recent research considers how to reasonably allocate resources for inference workloads~\cite{mohan2022looking, wang2021morphling} to optimize the system's resource utilization.


However, existing works have ignored two key opportunities to build a high-performance and cost-efficient model serving framework.
First, modern inference servers are equipped with general-purpose computing devices like CPUs and domain-specific computing devices like GPUs.
CPUs are usually used for input prepossessing, and GPUs for forward computation of the models~\cite{zhang2021charm} when handling deep learning inference workloads.
The increasing complexity of deep learning models and many inference requests result in higher GPU memory requirements and lower CPU resource utilization.
%
%
Hence, it is natural to consider collaboratively utilizing CPUs and GPUs for forward computation to improve system throughput and reduce inference latency instead of leaving CPUs idle and GPUs confined by limited memory capacity.

Second, the increasing complexity of deep learning models brings more potential to inter-operator parallelism during execution.
Multiple branches are implemented in each inception block of GoogleNet~\cite{szegedy2016rethinking}, one of the most classical convolutional neural networks. 
For recurrent neural networks (RNN), there are also opportunities for parallelism after unfolding along the number of layers and the sequence length.
Graph neural networks can also be executed in parallel due to their graph structure.
We find that RNNs are most suitable for hybrid execution on CPUs and GPUs because of their computation characteristics.
However, the current research on model parallel strategy mainly focuses on homogeneous devices~\cite{jia2019beyond, ding2021ios}, ignoring the potential of parallelism across general-purpose and domain-specific devices.

Therefore, we design \name, an inference framework designed for RNN models by collaboratively utilizing CPUs and GPUs for forward computation. 
The basic idea of \name is to exploit hardware capacity in the CPU-GPU environment, and its core component is a graph partition algorithm used to schedule RNNs' operators to heterogeneous platforms for parallel execution.
The system's inputs are a pre-trained RNN model in the ONNX format and the SLO requirement defined by the user.
\name first pre-processes the model, mapping the computation graph into a directed acyclic graph (DAG) and obtaining the execution time of operators both on CPUs and GPUs.
The graph partition module then sorts the operators in the directed acyclic graph and selects suitable execution providers for each operator. 
Finally, when users' inference requests arrive, \name will execute on CPUs and GPUs collaboratively according to the partition scheme and respond to the requests.

To confirm the effectiveness of our design, we conduct extensive experiments to evaluate our framework. We use the LSTM model as our baseline, the most widely used RNN model. 
By changing the parameters of the LSTM model, including the number of layers, input/output dimension, sequence length, and batch size, we generate more LSTM variants. 
Our experimental results show that, in the end-to-end experiment, the execution time of the model can be reduced by 19.4\% at most in the Latency-Optimal pattern compared with the execution on the GPU. 
By relaxing the latency limit of the model, our framework can reduce the GPU memory requirement of the model by 67.5\%. 
\name can reduce model swapping and SLO violation rate in the local cluster evaluation.

Our main contributions are as follows:

\begin{itemize}
    \item We identify the potential of parallel execution in the CPU-GPU environment for model inference. To our knowledge, we are the first to support inter-operator parallelism across heterogeneous platforms.
    \item We design \name, an optimized model serving framework for recurrent neural networks by collaboratively utilizing CPUs and GPUs to resolve GPU memory bottlenecks and improve CPU utilization.
    \item We design an adaptive graph partition algorithm to select execution platforms for operators. Our algorithm is not designed for specific models, so it can be extended to more complex models with little effort.
    \item We conduct extensive experiments on LSTM and its variants to confirm \name's high performance. Experimental results show that inter-operator parallelism in the CPU-GPU environment benefits RNN inference.
\end{itemize}

%% file: contents/02_background.tex
\section{Background and Motivation}



\begin{table}[!t]
    \centering
    \caption{GPU Memory Capacity \vs Inference Requirements.}
    \label{tab:memory}
    \resizebox{\linewidth}{!}{
        \begin{tabular}{@{}cc|cc@{}}
            \toprule
            \textbf{GPU}  & \textbf{Capacity} & \textbf{Model}       & \textbf{ Requirement}  \\ \midrule
            RTX 2080 Ti& 11 GB & VGG19~\cite{simonyan2014very}  & 1762 MB \\
            RTX 3090 Ti& 24 GB & YOLOv5~\cite{yolov5}   & 1444 MB\\
            Grid A100  & 48 GB & RenNext50~\cite{kone2018hierarchical} & 1218 MB\\
            \bottomrule
        \end{tabular}
    }
\end{table}

\begin{figure*}[!t]
    \begin{minipage}[c]{0.32\textwidth}
        \includegraphics[width=0.96\linewidth]{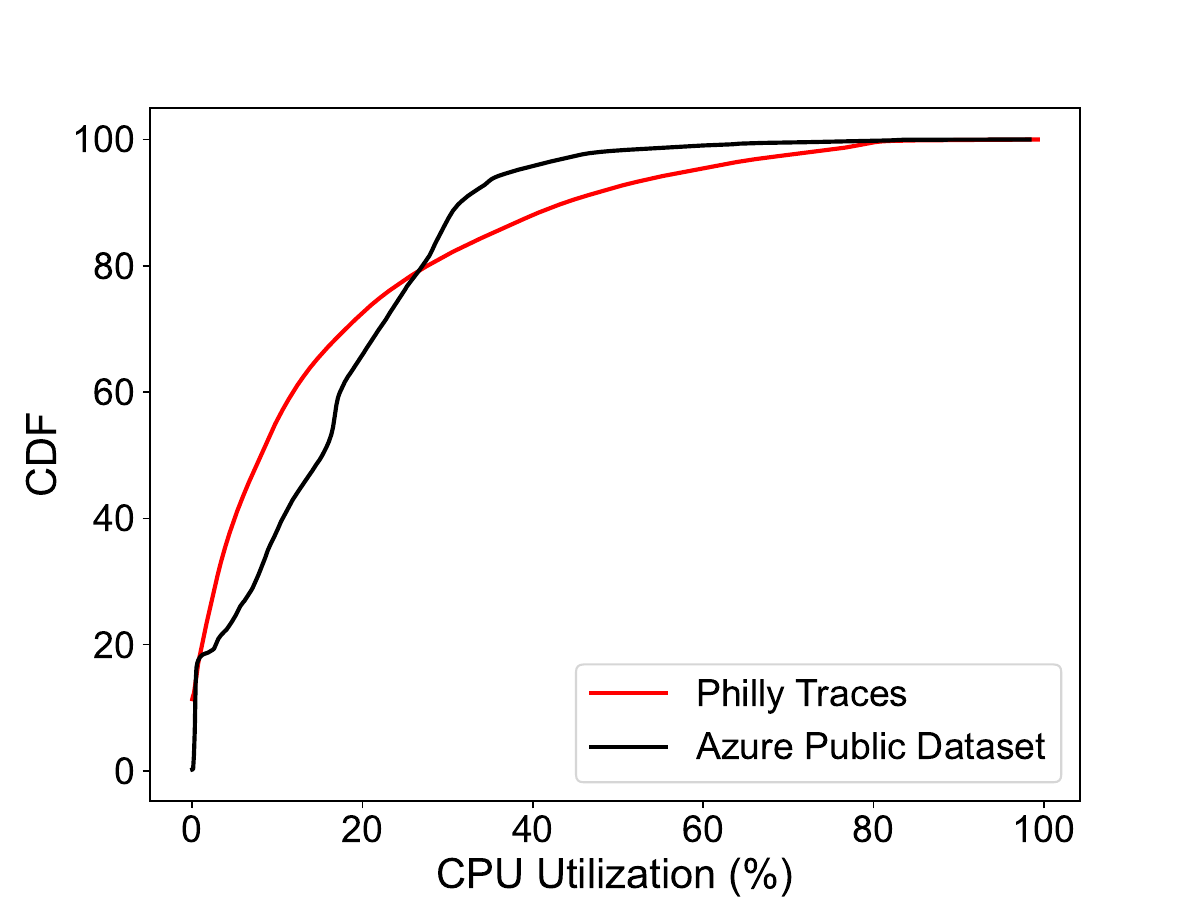}
        \caption{CPU Utilization in Clusters.}
        \label{fig:cpu_profiling}
    \end{minipage}
    \begin{minipage}[c]{0.32\textwidth}
        \centering
        \includegraphics[width=0.96\linewidth]{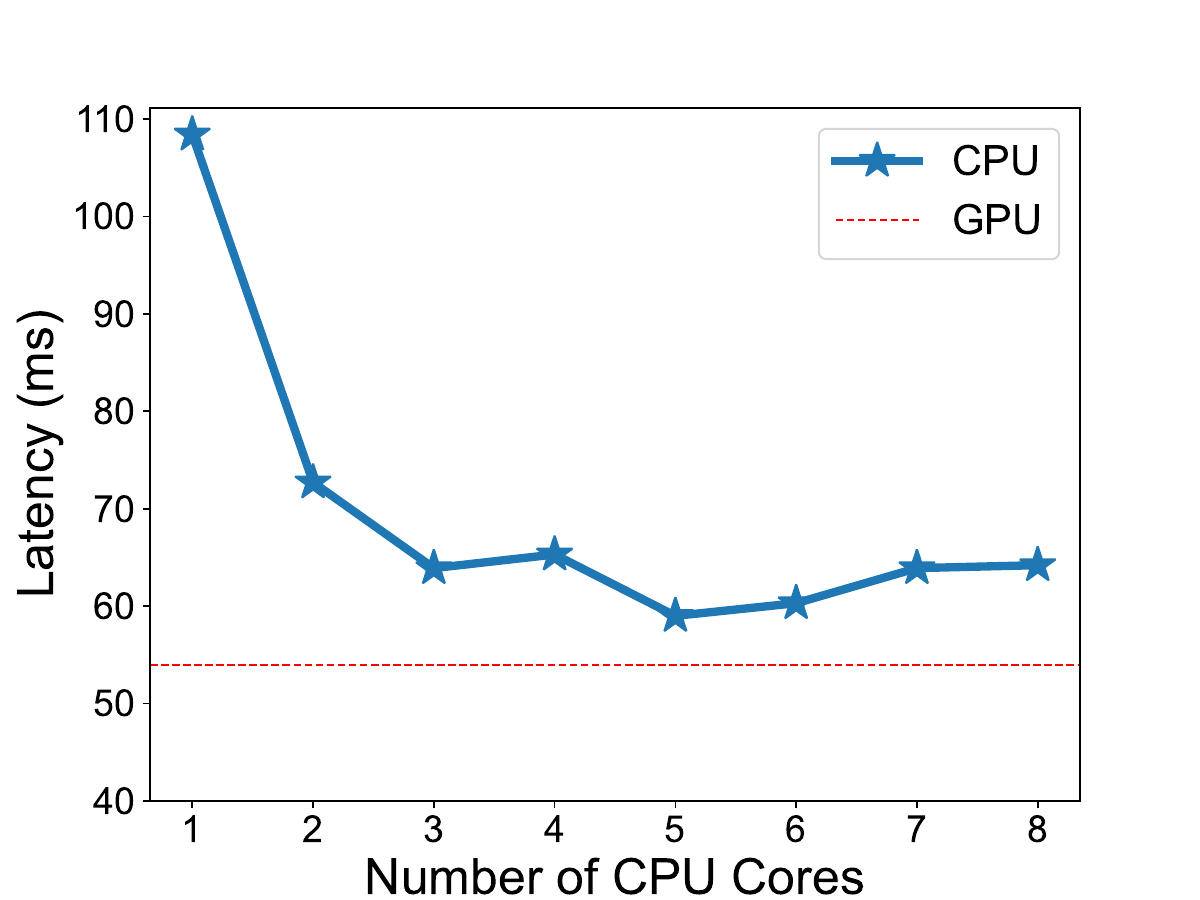}
        \caption{LSTM Profiling.}
        \label{fig:model_profiling}
    \end{minipage}
    \begin{minipage}[c]{0.32\textwidth}
        \centering
        \includegraphics[width=0.96\linewidth]{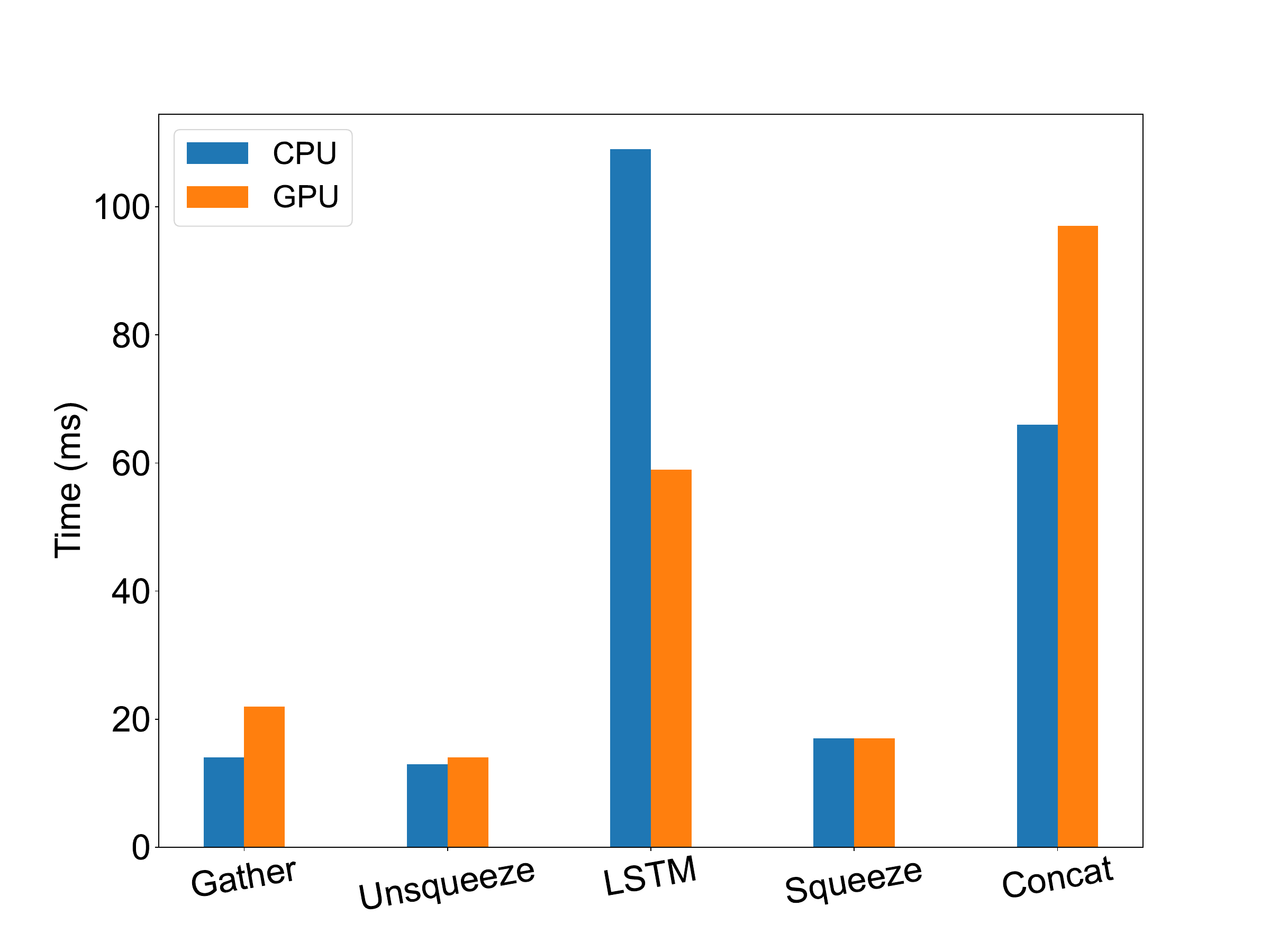}
        \caption{LSTM Operator Profiling.}
        \label{fig:op_profiling}
    \end{minipage}
\end{figure*}

\subsection{Serving Deep Learning Models}
Since deep learning plays an essential role in all walks of life, Inference-as-a-Service comes forth.
To build an efficient model reasoning system, researchers and engineers carry out optimization design from the following two aspects.
Since different DL models have different structures and characteristics, various techniques are proposed to optimize serving systems for specific model structures, like CNNs~\cite{liu2019optimizing, wei2018tgpa}, RNNs~\cite{gao2018low, zhang2018deepcpu, holmes2019grnn}, Transformers~\cite{fang2021turbotransformers, zhou2022pets, yu2022orca} and GNNs~\cite{mondal2022gnnie, liang2020engn}.
Others optimize the inference framework from the system perspective.
Morphling~\cite{wang2021morphling} provides an automatic resource configuration method for cloud-native model serving.
Clockwork~\cite{gujarati2020serving} serves as a distributed serving system with the observation of predictable model execution latency.
MArk~\cite{zhang2019mark} and Batch~\cite{ali2020batch} refer to the emerging cloud computing techniques to optimize inference efficiency.


\subsection{Observations and Opportunities}

\paragraph{Observation~\uppercase\expandafter{\romannumeral1}: GPU Memory Bottleneck in Deep Learning Clusters.} 
GPU memory becomes the bottleneck when serving thousands of inference requests in deep learning clusters.
Its consumption includes three parts when serving inference workloads: input/output tensors, weight tensors, and ephemeral tensors~\cite{gao2020estimating,gujarati2020serving}.
With the number of model parameters spiking from millions~\cite{krizhevsky2012imagenet, lecun1998gradient} to billions~\cite{brown2020language}, the memory requirement to load the whole model increases from MBs to GBs.
Ephemeral tensors are generated during invoking CUDA APIs, and GPU memory requirements also burst with the increasing number of GPU kernels.
Therefore, serving inference workloads with GPU has imposed intense pressure on GPU memory capacity, and Gao \etal conclude that 8.8\% of job failures in a deep learning cluster is caused by ``Out Of Memory''~\cite{gao2020estimating}.

However, GPU memory capacity is restricted compared with the increasing requirement of deep learning models.
\tab{memory} compares the memory capacity of mainstream NVIDIA GPUs \vs~the memory consumption of popular deep learning models when serving requests with a batch size equal to four.
We observe that only six VGG19 models can simultaneously be loaded in RTX 2080 Ti memory.
In contrast, a model serving cluster often serves requests with hundreds to thousands of models~\cite{soifer2019deep}. 
Therefore, when an inference request requires a model not in the GPU memory, we should first load the model from the main memory (much larger than GPU memory and hold all the models) and then perform forward computation.
Due to the limited GPU memory, model loading must be triggered frequently, significantly reducing the serving throughput and leading to violations of per-request deadlines. GPU memory is becoming even more critical to the serving performance as deep learning models get larger~\cite{gujarati2020serving}.

\vspace{4pt}
\begin{thmbox}[M]{\textbf{Takeaway~\uppercase\expandafter{\romannumeral1}}}
    \emph{Limited GPU memory capacity cannot satisfy the increasing demands of inference workloads, degrading the throughput of cluster servers.}
\end{thmbox}

\paragraph{Observation~\uppercase\expandafter{\romannumeral2}: Under-utilization of CPU Resources in the Cloud.}
The CPU utilization of deep learning clusters is typically low~\cite{cortez2017resource, jeon2019analysis}. We further verify this observation by analyzing the CPU utilization in modern cloud clusters from two open-source cloud datasets, Philly Traces~\cite{philly} and Azure Public Dataset~\cite{azure}.
The Philly Traces dataset collects information on clustered machines and task execution for Microsoft's machine learning platform during the four months in 2017, while the Azure Dataset collects task information for the cloud during 2019 and 2020.
The two curves in the \fig{cpu_profiling} can reflect the typical characteristics of CPU utilization in commercial clusters, regardless of whether for deep learning workloads.
About 80\% of CPUs in both Philly Traces and Azure Public Dataset are facing a utilization rate below 30\% while about 40\% of CPUs in Azure Public Dataset are below 40\%.

However, we observe CPU over-provisioning in commercial inference servers, which may aggravate the challenge of low CPU utilization.
Amazon EC2 G4 Instances~\cite{aws-g4} is claimed to be the most cost-effective GPU instances for machine learning inference in the industry.
We list resource configuration details of G4 Instances in \tab{aws_products} to illustrate the CPU over-provisioning status in commercial cloud platforms. 
The five configurations in Table~\ref{tab:aws_products} are equipped with a single GPU and multiple vGPUs.
When handling inference requests, only one CPU thread will be used to launch kernel functions if we only use GPU for forward computation, leaving most cores idle for scheduling.

\vspace{4pt}
\begin{thmbox}[M]{\textbf{Takeaway~\uppercase\expandafter{\romannumeral2}}}
    \emph{The waste of CPU resources in cloud clusters is severe, but they can become a supplement to the computing power on heterogeneous platforms.}
\end{thmbox}

\begin{table}[!t]
    \centering
    \caption{Resource Configuration of Amazon EC2 G4 Instances.}
    \label{tab:aws_products}
    \begin{tabular}{@{}lccc@{}}
        \toprule
        \textbf{Instance Size} & \textbf{GPU} & \textbf{vCPUs} & \textbf{Memory(GiB)} \\ \midrule
        \textbf{g4dn.xlarge} & 1 & 4 & 16 \\
        \textbf{g4dn.2xlarge} & 1 & 8 & 32 \\
        \textbf{g4dn.4xlarge} & 1 & 16 & 64 \\
        \textbf{g4dn.8xlarge} & 1 & 32 & 128 \\
        \textbf{g4dn.16xlarge} & 1 & 64 & 256 \\ \bottomrule
    \end{tabular}
\end{table}

\paragraph{Opportunity: Utilizing CPUs to Optimize RNN Inference.}

While GPUs are the primary choice to perform forward computation of deep learning models, CPUs are coming to the fore for some specific model structures.
Le \etal compare the speedups between an NVIDIA K80 GPU \vs~an Intel 20-core CPU for diverse deep learning models and reveals that CPUs and CPUs have similar performance when processing bi-directional LSTM models~\cite{le2020allox}.
We experiment with an 8-layer LSTM model, and \fig{model_profiling} and \ref{fig:op_profiling} demonstrates the overall model execution time and fine-grained operator execution time, respectively.

\begin{figure}[!t]
    \centering
    \subfloat[Folded RNN.\label{fig:folded-rnn}]{\includegraphics[width=0.12\textwidth]{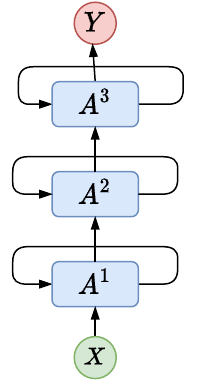}}\hfill
    \subfloat[Unfolded RNN.\label{fig:unfolded-rnn}] {\includegraphics[width=0.35\textwidth]{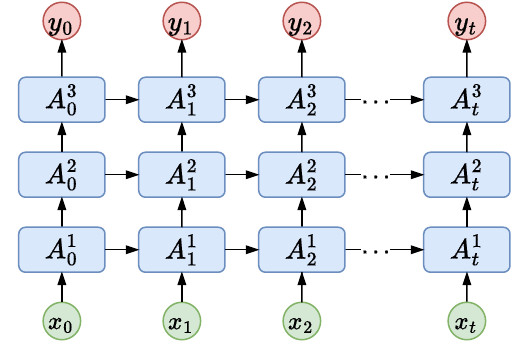}}\hfill
    \caption{Folded and unfolded RNNs. All the RNN cells are executed one-by-one when scheduled to one GPU, but they can exploit the multi-core parallelism of CPUs through inter-operator parallelism of data-independent kernels, \eg $A_0^2$ and $A_1^1$ can execute in parallel by schduling to different cores after the completion of $A_0^1$.}
    \label{fig:rnn}
\end{figure}

As a whole, we observe the change in the model's execution time with different CPU cores. In \fig{model_profiling}, as the number of CPU cores increases, the execution time of the model gradually decreases, approaching the execution time of the model executing on the GPU. 
But its execution time will fluctuate when the number of CPU cores is higher than five due to the limited parallel degree of model structure and contention for shared resources such as memory bandwidth, last level cache (LLC), etc.
We also analyzed the execution time of a single operator on the CPU and GPU, and the results are shown in \fig{op_profiling}. We find that not all GPUs have higher execution efficiency than the CPU. Moreover, the parallel ability of CPU multi-core can make up for the low execution efficiency of some operators on the CPU, as shown in \fig{rnn}.

\vspace{4pt}
\begin{thmbox}[M]{\textbf{Takeaway~\uppercase\expandafter{\romannumeral3}}}
    \emph{The multi-core parallelism of CPUs and the model structure of RNNs can be well-matched to optimize the inference performance of RNNs.}
\end{thmbox}

\subsection{Implications}


{\textbf{Takeaway~\uppercase\expandafter{\romannumeral1}}} and {\textbf{Takeaway~\uppercase\expandafter{\romannumeral2}}} imply that CPU can handle inference workloads in addition to GPUs in the deep learning cluster.
Introducing CPU to forward computation instead of handling all the workloads with GPU can both alleviate GPU pressure and improve the CPU utilization. Besides, compared with the limited GPU memory, the main memory in the deep learning cluster in ample, getting rid of OOM risk for the serving system.
Although the multi-core parallelism of the CPU effectively improves its task execution efficiency, there is still a gap in the performance of the CPU compared with the domain specific design of the GPU. 
Therefore, not all deep learning models are well suitable to adopt hybrid parallelism across heterogeneous platforms to optimize GPU memory footprint.
{\textbf{Takeaway~\uppercase\expandafter{\romannumeral3}}} indicates that RNNs are a suitable choice.

\begin{figure}[!t]
    \centering
    \subfloat[Demo Graph\label{fig:demo-graph}]{\includegraphics[width=0.2\textwidth]{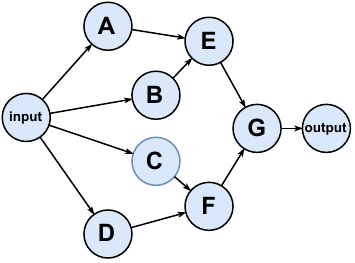}}\hfill
    \subfloat[Graph Partition.\label{fig:demo-partition}] {\includegraphics[width=0.25\textwidth]{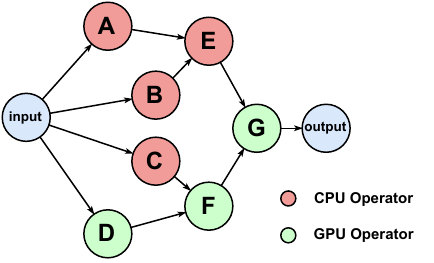}}\hfill
    \subfloat[Comparison of Three Execution Patterns.\label{fig:pattern-comparison}]{\includegraphics[width=0.45\textwidth]{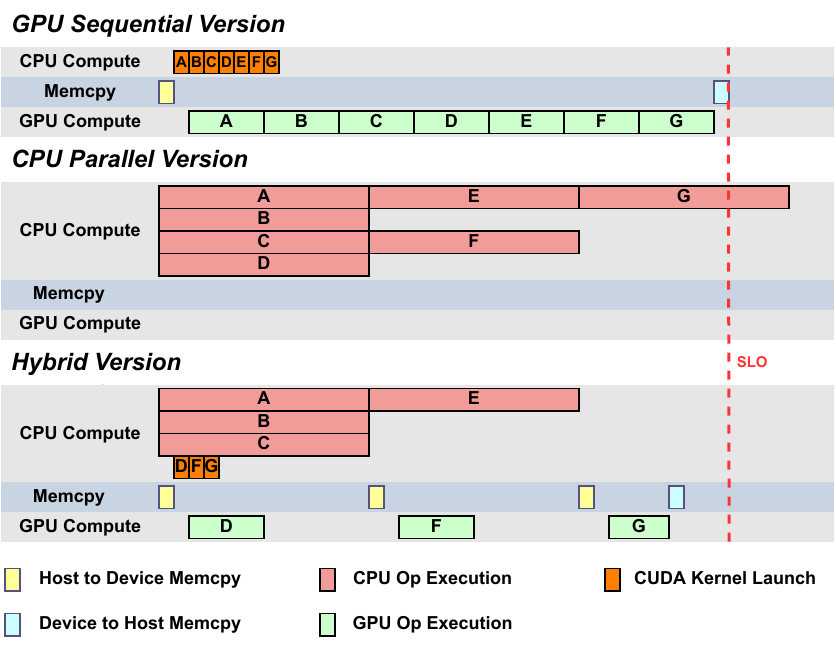}}
    \caption{Partition scheme for a seven-node computational graph. (a) and (b) illustrate the model structure and partition scheme, respectively. Four kernels are scheduled on GPU while the other three on CPU. (c) shows hybrid parallelism across CPUs and GPUs can minimize GPU memory requirement within SLO.}
    \label{fig:motivation}
\end{figure}

We demonstrate how to schedule different operators of a computing graph to heterogeneous platforms for efficient execution in \fig{motivation}.
Suppose there is a GPU graphics card and a 4-core CPU on a reasoning server, and the data copy between the CPU and the GPU is realized through PCIe. We need to schedule graph \textbf{G} with seven computing nodes to be executed in this heterogeneous environment.
It is the most common to schedule all the operators on GPU for efficient execution. In the GPU sequential version, we assume that the model weights have already been transferred to GPU memory. After the inputs are copied to GPU, kernel functions are launched on the CPU and executed on GPU sequentially. Finally, the outputs will be transferred back to the main memory and respond to users' requests to satisfy their SLO requirements.
Another method is to schedule the graph on the CPU when GPU memory is occupied. Unlike the sequential execution pattern, multi-branches models can utilize inter-operator parallelism to accelerate inference speed. For instance, we schedule \textbf{A}, \textbf{B}, \textbf{C}, \textbf{D} to different CPU cores in the first stage. Due to the data dependency of the computational graph, operator \textbf{G} can not be executed until the completion of \textbf{E} and \textbf{F}.

Since either CPU or GPU is fully utilized in the both execution patterns, we propose a hybrid execution pattern for multi-branch models in the heterogeneous environment.
We schedule the operator \textbf{A}, \textbf{B}, \textbf{C}, and \textbf{E} to be executed on CPU, and the others on CPU.
Since the four kernels from \textbf{A} to \textbf{D} are data-independent, they can run in parallel in heterogeneous platforms.
Operator \textbf{E} is scheduled on CPU to minimize data movement latency and save GPU time for operator \textbf{F}.
We can conclude that after computational graph partition and hybrid execution, our method can satisfy the users' SLO requirements and save GPU memory in the meantime.

\begin{figure}
    \centering
    \includegraphics[width=0.86\linewidth]{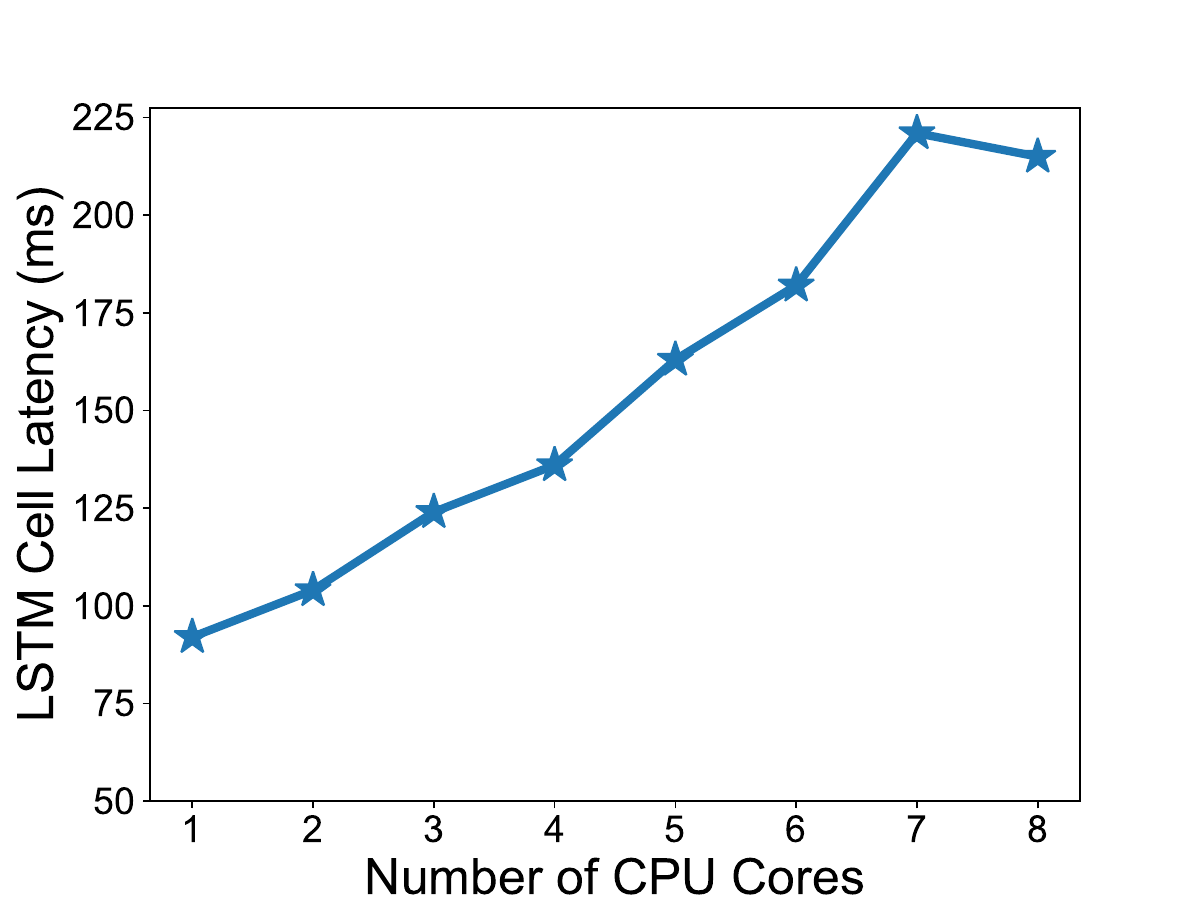}
    \caption{Cell Latency.}
    \label{fig:cell_latency}
\end{figure}

\subsection{Challenges}
The first challenge is efficiently partitioning a graph in fine granularity.   
Graph partition is commonly designed in machine learning scenarios. For instance, different graph partition algorithms are proposed for specific targets, such as overcoming memory bottleneck in serverless scenarios~\cite{yu2021gillis}, reducing context switch overhead by pipeline~\cite{bai2020pipeswitch}.
However, existing partition algorithms are designed either for homogeneous environments or models with a simple structure stacked sequentially of deep learning operators.
Layer grouping simplifies the problem formulation, which is unsuitable for much more complicated multi-branch models. 
To the best of our knowledge, \name is the first to provide a fine-grained graph partition for deep learning models to schedule the operators in a  heterogeneous environment. 

The second challenge is that the search space of the graph partition is ample due to the considerable number of model operators and variable hardware performance under different circumstances. 
The increase in the number of model operators makes the complexity of finding feasible scheduling schemes increase exponentially, which makes the time complexity of exhausting all feasible schemes and finding the best scheme unbearable. 
Besides, the selection of CPU cores also impacts performance. 
\fig{cell_latency} shows that the average latency of each LSTM cell increases when we use more CPUs for higher parallelism, which results from shared resource contention, including memory bandwidth and LLC.

%% file: contents/03_formulation.tex
\section{Problem Formulation}

\paragraph{Problem Description}
We map the computational graph of a machine learning model to a directed acyclic graph $G = (V, E)$.
The vertex set $V = \{v_0, v_1, \cdots, v_{n-1}\}$ represents the $n$ kernel functions in the computational graph.
The edge set $E = \{e_{ij} | 0 \le i \neq j < n\}$ represents the execution dependency between kernel functions, where $e_{ij}$ represents that fuction $v_j$ should be launched only after the execution of function $v_i$.  
Based on the mapping of the computational graph, we define the predecessor and successor sets for each vertex $v_i$: $\operatorname{pred}(v_i) = \{v_j | e_{ji} \in E, j \in [0, n)\}$ and $\operatorname{succ}(v_i) = \{v_j | e_{ij} \in E, j \in [0, n)\}$.
The vertex that has an empty predecessor set is the entry function of the graph, while the vertex that has an empty successor set is the exit function. Note that neither the entry function nor the exit function is unique in the computational graph.

We denote available processors as $P = \{p_0, p_1, \cdots, p_{k}\}$, where $p_0$ represents a GPU, and there are $k$ available CPU cores in the heterogeneous environment.
Although we assume that all the $k$ CPU cores have the same computation capacity, they have diverse performance when different numbers of CPU cores are in use. 
We define a performance weight matrix $W \in \mathbb{R}^{n \times (k+1)}$ to record the execution latency. $W_{i,0}$ denotes how long to execute function $v_i$ on GPU, while $W_{i, j} (j \neq 0)$ denotes how long to execute function $v_i$ on CPU when $j$ CPU cores are in use.
Since the communication cost between CPU and GPU is not negligible in the heterogeneous environment, we define $C \in \mathbb{R}^{n \times n}$ as the communication data size and $b$ as the bandwidth between CPU and GPU. Note that $C_{i, j} = 0$ if $e_{ij} \notin E$ and we ignore the communication cost between $v_i$ and $v_j$ if they are scheduled on the same device since the memory bandwidth is much larger than PCIe bandwidth between heterogeneous platforms.
$M$ is a memory consumption matrix whose dimension is $n \times 4$. $M_{i, 0}, M_{i, 1}, M_{i, 2}, M_{i, 3}$ denote the GPU memory requirements of input tensors, output tensors,  ephemeral tensors and model weights, respectively, if $v_i$ is scheduled on GPU.

\begin{table}[!t]
\centering
\caption{Notation Table.}
\begin{tabular}{cp{2.5in}} 
\toprule
Notation & \multicolumn{1}{c}{Description}  \\ 
\midrule
$G$ & a computational graph          \\
$V$ & a set of kernel functions $\{v_0, v_2, \cdots, v_{n-1}\}$ \\
$E$ & execution dependency among vertex \\
$\operatorname{pred}(v_i)$ & immediate predecessors of vertex $v_i$ \\
$\operatorname{succ}(v_i)$ & immediate successors of vertex $v_i$ \\
$P$ & a list of available processors \\
$W$ & a $n \times (k+1)$ matrix of performance \\
$C$ & a $n \times n$ matrix of communication data size\\
$M$ & a $n \times 4$ matrix of memory consumption \\
$n$ & number of vertex in the graph \\
$k$ & number of available CPU cores \\
$b$ & PCIe bandwidth between CPU and GPU \\
\midrule
$\operatorname{EST}(v_i, p_j)$ & earliest start time of $v_i$ if scheduled on $p_j$\\
$\operatorname{EFT}(v_i, p_j)$ & earliest finish time of $v_i$ if scheduled on $p_j$\\
$\operatorname{AFT}(v_i)$ & actual finish time of $v_i$\\
\midrule
$O$ & execution order\\
$S$ & device selection\\
$k^*$ & selected number of CPU cores\\
\midrule
$L$ & execution latency\\
$M$ & GPU memory requirement\\
\bottomrule
\end{tabular}
\end{table}

\paragraph{Objective \& Constraints}
Our problem considers how to generate a partition plan to schedule the computational graph $G$ in a CPU-GPU heterogeneous environment.
The expected outputs should minimize the GPU memory consumption within the guarantee of inference latency.
We formulate the objective as Equation \ref{eq:target}, where $\alpha$ is a hyper-parameter to make trade-off between inference delay and memory consumption. The execution latency $L$ of the computation graph is equal to the maximum actual finish time of all the exit functions considering multiple exit functions in the graph: 
\begin{equation}
    L = \max_{\operatorname{succ}(v_i) = \varnothing} \left\{\operatorname{AFT}(v_i)\right\}
\end{equation}

\noindent \eq{est} and \eq{eft} illustrates how to compute $\operatorname{EST}(v_i, p_j)$ and $\operatorname{EFT}(v_i, p_j)$ if the function $v_i$ is scheduled on $p_j$. $\operatorname{EFT}(v_i, p_j)$ is equal to the sum of $\operatorname{EST}(v_i, p_j)$ and its execution time on processor $p_j$. If GPU is selected, the execution time equals $W_{i, 0}$ while it equals $W_{i, k^{\prime}}$ since we assume the CPU capacity is bottle-necked by the number of running CPU cores in the heterogeneous environment. 
\begin{equation}\label{eq:est}
    \operatorname{EST}\left(v_{i}, p_{j}\right)=\max \left\{\operatorname{avail}[j], \\
    \max_{v_{m} \in \operatorname{pred}\left(v_{i}\right)}\left(\operatorname{AFT}\left(v_{m}\right)+\frac{C_{m,i}}{b}\right)\right\}
\end{equation}

\begin{equation}\label{eq:eft}
\operatorname{EFT}\left(v_{i}, p_{j}\right)=
\begin{cases}
W_{i, 0}+\operatorname{EST}\left(v_{i}, p_{j}\right), \text{ if } j = 0 \\
W_{i, k^{\prime}}+\operatorname{EST}\left(v_{i}, p_{j}\right), \text{ if } j \neq 0
\end{cases}
\end{equation}

\noindent We formulate the memory consumption of the computational graph as:

\begin{equation}\label{eq:mem}
    M = \sum_{v_i \in V}(s_i==0) \cdot (M_{i, 1} + M_{i, 2} + M_{i, 3} + \sum_{j \in \operatorname{pred}(i), s_j \neq 0}M_{j, 1})  
\end{equation}

The partition plan consists of three components: an execution order $O$, a device selection $S$, and the number of CPU cores $k^*$. 
Constraints \ref{eq:order-1} and \ref{eq:order-2}  ensure that the execution order $O = \{o_0, o_1, \cdots, o_{n-1}\}$ is some kind of permutation of integers from $0$ to $n-1$. 
Constraint \ref{eq:order-3} means that the execution order should satisfy the requirements of the topological sorting since a child node can only be executed after its predecessors.
The device selection $S = \{s_1, s_2, \cdots, s_n\}$ denotes whether to launch the kernel function on the GPU or CPU, and if $s_i$ equals 0, the vertex $v_i$ is scheduled to be executed on GPU (Constraint \ref{eq:s}).
The selected number of CPU cores $k^*$ is an integer between $0$ and $k$ (Constraint \ref{eq:k}). If $k^*$ is set to $0$, the computational graph is scheduled only on GPU without optimization of GPU memory consumption.

In all, our problem can be formulated as the optimization problem below:

\noindent \textbf{minimize}
\begin{equation}\label{eq:target}
   L + \alpha M
\end{equation}

\noindent \textbf{subject to}
\begin{align}
    o_i \in [0, n)&, \quad \forall i \in [0, n) \label{eq:order-1} \\
    o_i \neq o_j&, \quad \forall i, j \in [0, n) \text{ and } i \neq j \label{eq:order-2} \\
    o_i < o_j&, \quad v_j \in \operatorname{succ}(v_i) \label{eq:order-3} \\
    s_i \in \{0, 1\}&, \quad \forall i \in [0, n) \label{eq:s}\\
    0 \le k^* \le k \label{eq:k} & \\
    \operatorname{EST}\left(v_i, p_{j}\right)=0&, \quad \operatorname{pred}(v_i) = \varnothing
\end{align}

%% file: contents/04_design.tex
\section{System Design}

\input{contents/4/4_1_architecture}

\input{contents/4/4_3_graph_partition.tex}

%% file: contents/4/4_1_architecture.tex
\subsection{System Architecture}

\paragraph{Design Goal.} \name's design satisfies three requirements for a machine learning serving system.
Firstly, \name is portable for machine learning developers because its inputs are only pre-trained models and user-defined SLO requirements.
It does not require developers to spare more effort to transform their models to a specific format since ONNX is supported by the most popular deep learning frameworks, including TensorFlow~\cite{tensorflow}, PyTorch~\cite{pytorch} and MXNet~\cite{mxnet}.
Secondly, \name is lightweight to satisfy the urgent latency requirements of inference requests. 
Existing methods resort to more complex algorithms to decide configurations to serve machine learning inference workloads, like meta-learning~\cite{wang2021morphling}, LSTM~\cite{zhang2019mark}, which has been a drag on the prompt response to users' inference requests. 
\name utilizes a classical graph traversing algorithm for topological sorting and a greedy algorithm for device selection, which are much more light-weighted than ML algorithms.
Finally, \name provides a universal schedule plan independent of model structures for computational graph partition. \name is not elaborately designed for one specific model but can be extended to much more complex multi-branch models.

\begin{figure}[!t]
    \centering
    \includegraphics[width=0.50\textwidth]{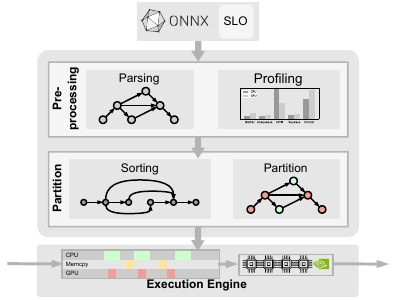}
    \caption{System Architecture.}
    \label{fig:system_overview}
\end{figure}

\paragraph{Workflow.} After training the deep learning models, developers can submit their models in the ONNX format and define the expected execution latency. 
Then \name will derive a directed acyclic graph of the model and profile the model to get the operator execution time on heterogeneous platforms. 
The DAG is significant because it not only determines the execution order of the model's operators but also affects the memory copy between host and device memory. 
The data movement happens when two adjacent operators are scheduled to different devices.
Graph partition is the core component of \name, composed of two algorithms, topological sorting and device selection. The former algorithm decides the execution order of the computational graph, and the latter chooses which device is most suitable for each operator.
When the user sends a request for model inference, the runtime will check whether the user's model weight has been copied to the GPU's display memory. 
When the model is not in the GPU memory, and the remaining memory can not meet the minimum requirements for executing the calculation, the runtime will select a replacement algorithm to cache the model in the CPU. 
Finally, \name will use the CPU and GPU to execute the model operators in parallel and return the execution results within the requirements of the SLO.

%% file: contents/4/4_3_graph_partition.tex
\subsection{Topological Sorting}

\begin{figure}[!h]
    \centering
    \includegraphics[width=0.45\textwidth]{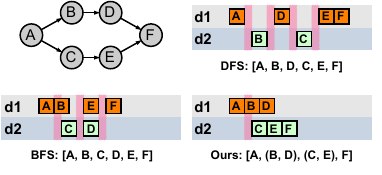}
    \caption{Topological Sorting Optimization.}
    \label{fig:sorting-optim}
\end{figure}

Given a computational graph, it is critial to determine the execution order for the vertices on the graph, which influences the inference efficiency on heterogeneous platforms and SLO satisfaction rate for users.
Depth-first-search (DFS) algorithm and breadth-first-search (BFS) algorithm are the two most commonly used graph traversal algorithms, which have been widely used in topological sorting.
Both starting from the root nodes, DFS explores feasible kernel functions as far as possible before backtracing, while BFS explores vertices by the order of their distances from the root nodes.
However, \fig{sorting-optim} demonstrates that simply applying BFS or DFS can not achieve ideal benefits. 
We assume there are two available devices and we always schedule the nodes according to the given order to a more idle device. 
It results in low parallelism if scheduling the kernel functions according to DFS, while high context switches between continuous kernels if according to BFS. 
The optimal situation should be that we can explore parallelism with BFS and maintain locality to reduce context switch with DFS.

\begin{algorithm}[!t]
    \caption{Topological Sorting}
    \label{alg:sorting}
    \SetKwData{In}{\textbf{in}}
    \SetKwData{To}{to}
    \DontPrintSemicolon
    \SetAlgoLined
    \KwIn {computational graph $G = (V, E)$}
    \KwOut {schedule order $O$}
      \Begin{
          initialize an empty order $O$ \;
          initialize an empty ready queue $Q$ \;
          initialize a mark list $marked \gets [false] \times n$ \;
          \For{$v_i$ in $V$}{
                \If{$\operatorname{pred}(v_i) = \varnothing$}{
                    push $v_i$ into $Q$ \;
                }
          }
          \While{$Q$ is not empty}{
              pop the first element $v_{curr}$ from $Q$
              $ready = true$ \;
              $ready = ready \enspace  \&\& \enspace marked[v_j], v_j \in \operatorname{pred}(v_{curr})$ \;
              \eIf{$ready$}{
                  $marked[v_{curr}] = true$ \;
                  append $v_{curr}$ to $O$ \;
              }{
                push $v_{curr}$ into $Q$ \;
              }
          }
        \Return{$O$}
    }
\end{algorithm}


Based on the observations above, we design a topological sorting method by taking advantage of BFS and DFS, and provide the pseudo code in Algorithm \ref{alg:sorting}.
We use BFS to expolore potential parallelism by distributing different branches to different devices for execution, and DFS to maintain locality by avoiding closely-related neighbour kernel functions being scheduled across devices.
In Algorithm \ref{alg:sorting}, we first initialize three variables, order $O$ to be returned by the algorithm, a ready queue $Q$ to store the those kernels that have no predcessors or whose kernels have been scheduled, and a mark list to record whether a kernel has been scheduled.
Different from the classical BFS algorithm, before appending the ready kernel $v_{curr}$ to $O$, we will check whether its child function can be merged.
If its child has only one predcessor and the communication time is larger than the average execution time among devices, it means maintaining the locality can achieve more benefits, and it is where DFS plays its role in the algorithm. 

\subsection{Device Selection}

\begin{algorithm}[!t]
    \caption{Device Selection}
    \label{alg:selection}
    \SetKwData{In}{\textbf{in}}
    \SetKwData{To}{to}
    \DontPrintSemicolon
    \SetAlgoLined
    \KwIn {computational graph $G = (V, E)$, topological order $O$}
    \KwOut {device selection $S$, selected number of CPU cores $k^{*}$}
        \Begin{
        $\operatorname{TotalCosts} \gets []$ \;
        \For{$k^{\prime}$ in $[0, k]$}{
            $S_{k^{\prime}} \gets []$ \;
            $\operatorname{avail}[0, \cdots, k^{\prime}] \gets [0, \cdots, 0]$ \;
            \For{$i$ in $O$} {
                $\operatorname{Costs} \gets []$ \;
                \For{$j$ in $[0, k^{\prime}]$} {
                    compute $\operatorname{EST}(v_i, p_j)$ with Eq. \ref{eq:est} \;
                    compute $\operatorname{EFT}(v_i, p_j)$ with Eq. \ref{eq:eft} \;
                    compute $M(v_i, p_j)$ with Eq. \ref{eq:mem} \;
                    $\operatorname{cost}_j = \operatorname{EFT}(v_i, p_j) + \alpha{M(v_i, p_j)}$ \;
                    add $\operatorname{cost}_j$ to $\operatorname{Costs}$
                }
                set $S_{k^{\prime}}[i]$ to $j$ with the minimum $\operatorname{Costs}[j]$ \;
                update $L$ and $M$\;
            }
            add $(L + \alpha{M})$ to $\operatorname{TotalCosts}$ \;
        }
        select $k^{*}$ with the minimum $\operatorname{TotalCosts}[k^{*}]$ \;
        set device selection $S = S_{k^{*}}$ \;
        \Return{$k^{*}$, $S$} \;
    }
\end{algorithm}

The increasing number of model operators and variable hardware performance make it impractical to enumerate all the device selections to achieve the optimal. 
We design an adaptive algorithm to provide fine-grained device selection for each operator in the computational graph. 
The core idea of our algorithm is to use greedy method to select the appropriate device for each operator to perform the forward computation. 
After comparing the execution time of the operator with the available time of the hardware, the greedy algorithm always schedules the operator to the device that can let the operator end the execution earliest.

We show the details of the device selection algorithm in Algorithm \ref{alg:selection}. 
The inputs of the adaptive device selection are the DAG and the topological order obtained by Algorithm \ref{alg:sorting} while its outputs are the optimal resource configuration and the device selection scheme for each operator.
\name computes the cost for each kind of resource configuration. When the number of chosen CPU cores equals $k^\prime$, we first initialize the available time for each device (Line 5). 
Then we iterate over the topological order list to find the best device for each operator (Line 8-14).
For each operator $o_i$ in the order $O$, \name calculate the earliest finish time and memory requirement, and then decide which device to select to achieve the least cost.



%% file: contents/impl.tex
\section{Implementation}

\begin{figure}[!t]
    \centering
    \includegraphics[width=0.90\linewidth]{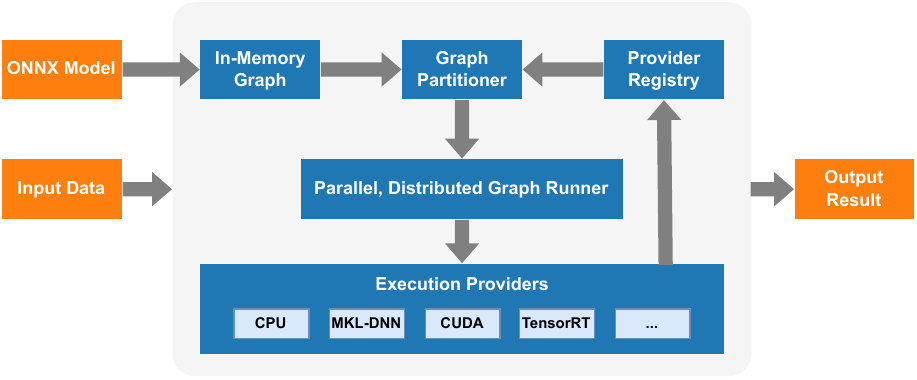}
    \caption{ONNXRuntime Architecture.}
    \label{fig:onnxruntime}
\end{figure}

\name is implemented based on ONNXRuntime~\cite{onnxruntime}, a cross-platform inference and training accelerator for machine learning workloads.
\fig{onnxruntime} shows the high-level design of ONNXRuntime.
It first converts users' input graph into an in-memory graph presentation, and performs graph optimizations independent of execution providers like operator fusing and constant folding.
The execution provides are the intersection of user-defined and system-provided execution providers.
The graph partitioner splits the graph into sub-graphs and schedule them to execute on different devices.
ONNXRuntime implements a simple partition technique that schedules the graph according to the order of user-defined providers.  
Finally, the parallel and distributed graph runner execute the sub-graphs after accepting uses' input data on the underlying devices, like CPU and CUDA.

Our modification on ONNXRuntime focuses on the following.
Firstly, based on the default in-memory graph, we implement the graph parsing and profiling component. These two components are independent of ONNXRuntime for flexibility.
Secondly, we re-implement the graph partition algorithm. Different from the default graph partition according to the user-defined order, \name provides a dynamic graph partition algorithm to allow parallel execution between CPUs and GPUs.
Finally, we re-design the execution mechanism of providers. Since the default execution providers has no distinction between CPU and GPU operators, we add two additional operators, GPU operator and memory-copy operator and assign these operators to different thread pools for execution.

%% file: contents/05_evaluation.tex
\section{Evaluation}

\input{contents/5/5_1_setup.tex}

\input{contents/5/5_2_e2e.tex}

\input{contents/5/5_3_component.tex}

\input{contents/5/5_4_cluster.tex}

%% file: contents/5/5_1_setup.tex
\subsection{Experiment Setup}

We evaluate \name to reveal its high-performance in terms of latency and memory requirement when serving recurrent neural networks.
\tab{hardware} and \tab{software} shows the hardware and software requirements of our experiments.
In the rest of the section, our experiments are designed to answer the following questions.
\begin{itemize}
    \item \textbf{How can \name optimize response time or memory requiremnet for inference workloads?} The end-to-end experiments in \S\ref{sec:e2e} show \name can reduce inference latency by 13.3\% in the latency-optimal scenario and reduce GPU memory requirement by 60.0\% in the memory-optimal scenario.
    \item \textbf{How does the hyperparameter affect graph partition?} The component analysis of $\alpha$ in \S\ref{sec:alpha} shows that we can adjust $\alpha$ to achieve trade-off between GPU memory requirement and inference latency.
    \item \textbf{How can \name performance for real workloads in local clusters?} We test \name with real-world workloads in \S\ref{sec:local} and it can reduce GPU memory loading and offloading by from 44\% to 9\%.
\end{itemize}

\begin{table}[!t]
    \centering
    \caption{Hardware Proterties for Experiments.\label{tab:hardware}}
    \resizebox{\linewidth}{!}{%
        \begin{tabular}{@{}cc|cc@{}}
        \toprule
        \textbf{Component} & \textbf{Specification} & \textbf{Component} & \textbf{Specification} \\ \midrule
        CPU Device & Intel Xeon E5-2685 v3 & GPU Device & NVIDIA RTX 1080Ti \\
        Memory Capacity & 128GB & GPU Memory & 12GB \\
        Number of Cores & 24 (12 physical cores) & GPU SM Cores & 4352 \\
        Shared LLC Size & 30MB & Operating System & Ubuntu 18.04 \\ \bottomrule
    \end{tabular}%
    }
\end{table}

\begin{table}[!t]
    \centering
    \makeatletter\def\@captype{table}\makeatother\caption{Software Proterties for Experiments.\label{tab:software}}
    \resizebox{\linewidth}{!}{%
    \begin{tabular}{@{}cc|cc@{}}
        \toprule
        \textbf{Component} & \textbf{Specification} & \textbf{Component} & \textbf{Specification} \\ \midrule
        ONNXRuntime & v1.10.0 & onnx & v1.12.0 \\
        CUDA & 11.0 & cudnn & 8 \\
        Python & 3.8 & & \\\bottomrule
    \end{tabular}
    }
\end{table}

%% file: contents/5/5_2_e2e.tex
\subsection{End-to-End Experiments\label{sec:e2e}}


\begin{table*}[!h]
    \centering
    \caption{End-to-End Experiments. The structure of the baseline model is $\mathtt{num\_of\_layers}=12, \mathtt{batch\_size}=8, \mathtt{io\_size}=64, \mathtt{seq\_len}=96$. We adjust only one structural parameter for other models, and compare their performance in four scenarios: GPU, CPU, Latency-Optimal, and Memory-Optimal. ``--'' means \name is not feasible to optimize such a model structure.}
    \label{tab:e2e}
    \resizebox{\textwidth}{!}{
    \begin{tabular}{@{}ccccccccccccccc@{}}
    \toprule
    \multirow{2}{*}{\textbf{Pattern}} & \multirow{2}{*}{\textbf{Metric}} & \multirow{2}{*}{\textbf{Baseline}} & \multicolumn{3}{c}{\textbf{Number of Layers}} & \multicolumn{4}{c}{\textbf{Batch Size}} & \multicolumn{2}{c}{\textbf{I/O Size}} & \multicolumn{3}{c}{\textbf{Sequence}} \\ \cmidrule(l){4-6} \cmidrule(l){7-10} \cmidrule(l){11-12} \cmidrule(l){13-15} 
     &  &  & \textbf{4} & \textbf{8} & \textbf{16} & \textbf{1} & \textbf{2} & \textbf{4} & \textbf{16} & \textbf{32} & \textbf{128} & \textbf{32} & \textbf{64} & \textbf{128} \\ \midrule
    \multirow{2}{*}{\textbf{GPU}} & GPU Memory (MB) & 1643.0 & 777.0 & 1209.0 & 2075.0 & 1643.0 & 1643.0 & 1643.0 & 1643.0 & 1533.0 & 2461.0 & 777.0 & 1209.0 & 2075.0 \\
     & Latency (ms) &  59.3 & 20.5 & 39.0 & 78.1 & 58.3 & 59.1 & 57.2 & 58.2 & 59.2 & 58.6 & 19.7 & 39.2 & 78.5 \\ 
     \midrule
    \textbf{CPU} & Latency (ms) & 70.6 & 23.9 & 47.1 & 94.1 & 65.9 & 63.3 & 67.0 & 80.4 & 61.5 & 96.8 & 23.7 & 47.1 & 95.4 \\\midrule
    \multirow{4}{*}{\textbf{\begin{tabular}[c]{@{}l@{}}Latency-\\ Optimal\end{tabular}}} & GPU Memory (MB) & 948.1 & 525.0 & 833.0 & 1189.0 & 959.0 & 935.0 & 1053.0 & 623.0 & 927.0 & 875.0 & 609.0 & 757.0 & 707.0 \\
    & Memory Reduction (\%) & 42.3 & 32.4 & 31.1 & 42.7 & 41.6 & 43.1 & 35.9 & 62.1 & 39.5 & 64.4 & 21.6 & 37.4 & 65.9 \\
     & Latency (ms) & 52.0 & 21.2 & 36.8 & 71.1 & 48.2 & 47.6 & 55.7 & 67.5 & 52.4 & 72.0 & 18.8 & 34.2 & 82.1 \\ 
     & Latency Reduction(\%) & 12.3 & -3.2 & 5.6 & 8.9 & 17.4 & \textbf{19.4} & 2.7 & -16.0 & 11.6 & -22.9 & 4.2 & 12.7 & -4.5 \\ \midrule
    \multirow{4}{*}{\textbf{\begin{tabular}[c]{@{}l@{}}Memory-\\ Optimal\end{tabular}}} & GPU Memory (MB) &  629.0 & --- & 833.0 & 675.0 & 607.0 & 591.0 & 1053.0 & --- & 581.0 & --- & 609.0 & 533.0 & ---  \\
    & Memory Reduction (\%) & 61.7 & --- & 31.1 & \textbf{67.5} & 63.1 & 64.0 & 35.9 & --- & 62.1 & --- & 21.6 & 55.9 & --- \\
     & Latency (ms) & 58.4 & --- & 36.8 & 77.3 & 57.2 & 55.8 & 55.7 & --- & 55.8 & --- & 18.8 & 39.1 & ---  \\ 
     & Latency Reduction (\%) & 1.6 & --- & 5.6 & 1.0 & 1.9 & 5.7 & 2.7 & --- & 5.7 & --- & 4.2 & 0.4 & --- \\ \bottomrule
    \end{tabular}}
\end{table*}

\begin{figure*}[!t]
    \centering
    \subfloat[$\langle{12, 8, 64, 96}\rangle$.\label{fig:alpha_baseline}]{\includegraphics[width=0.24\linewidth]{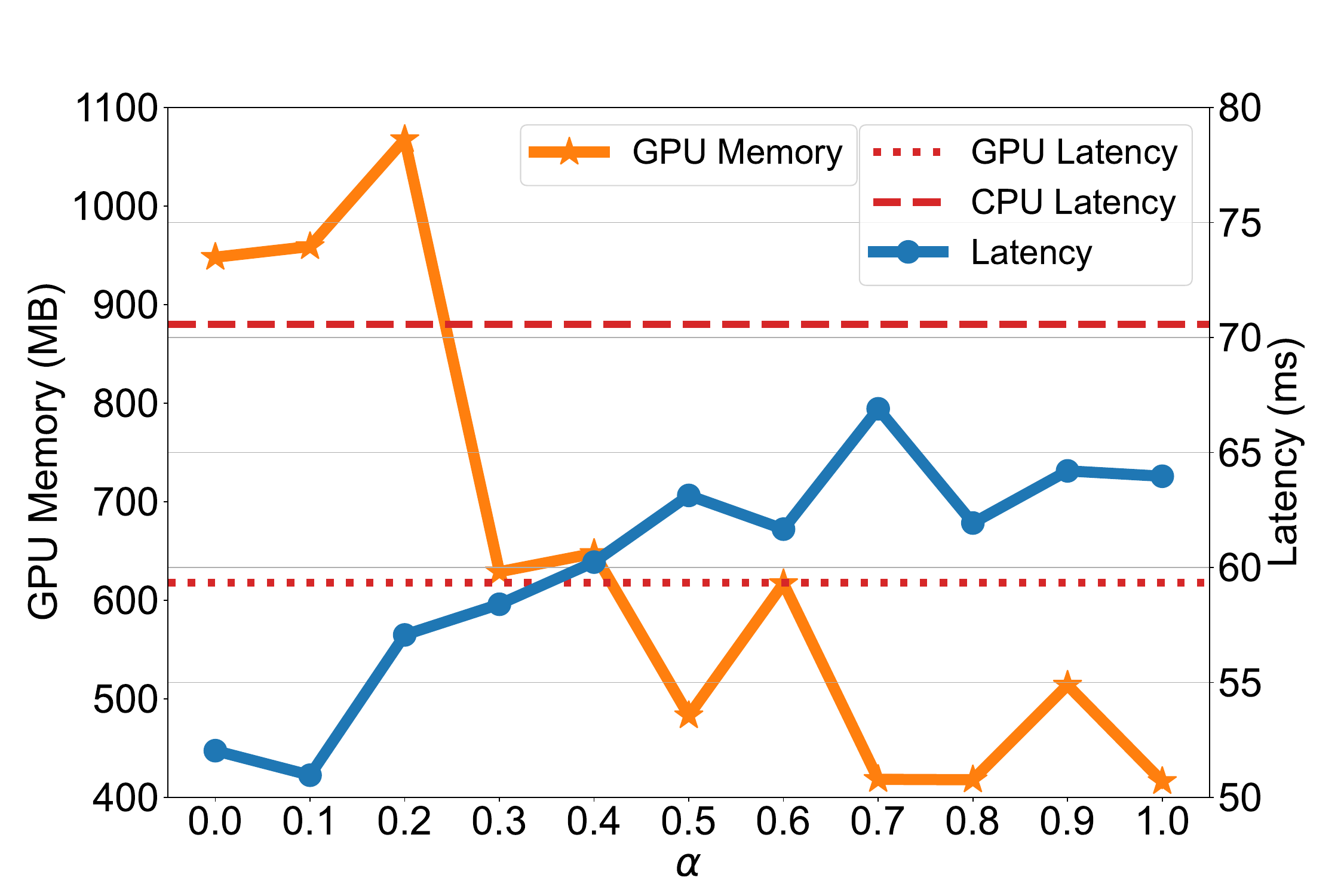}}\hfill
    \subfloat[$\langle{12, 1, 64, 96}\rangle$.\label{fig:alpha_bs}] {\includegraphics[width=0.24\linewidth]{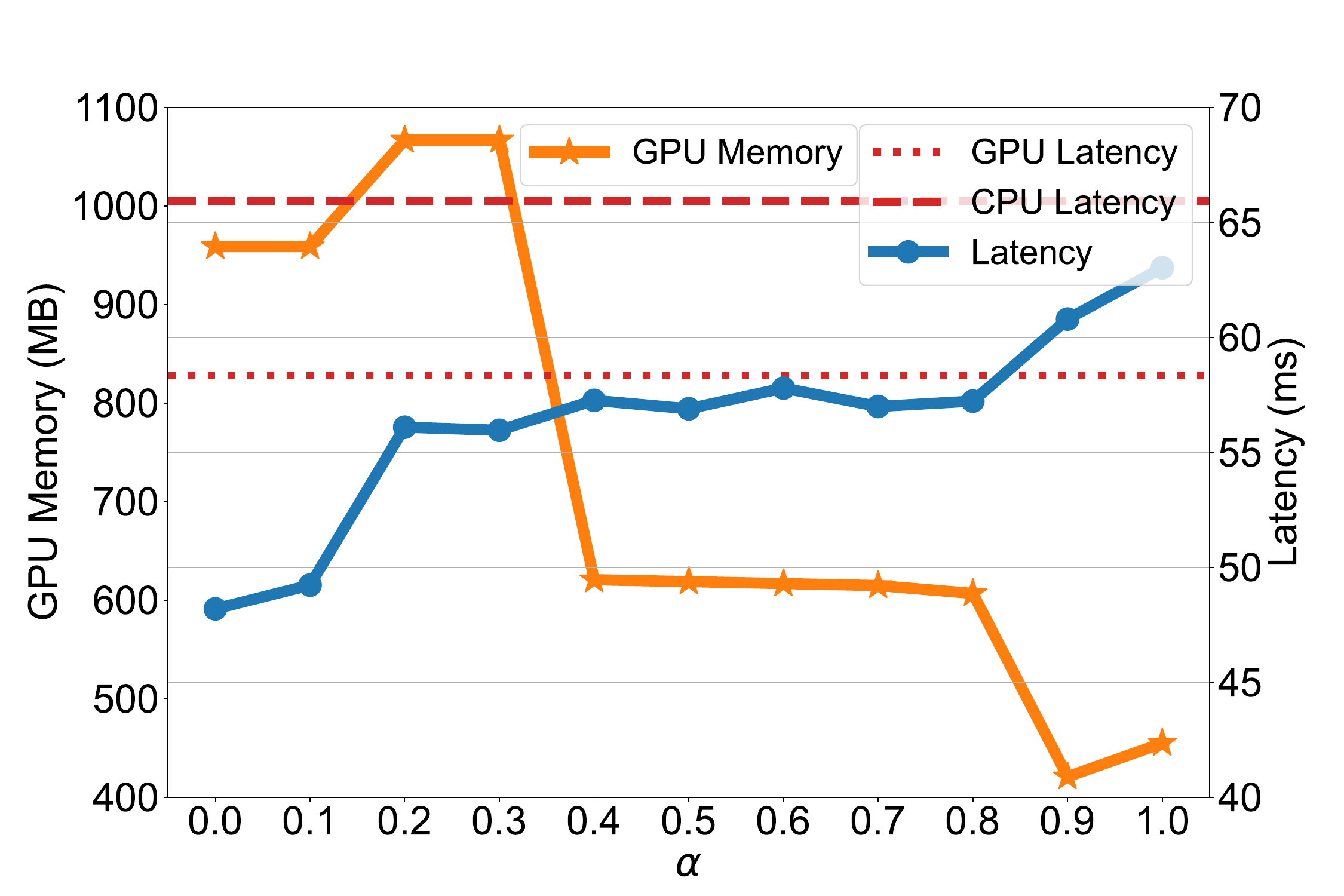}}\hfill
    \subfloat[$\langle{12, 8, 32, 96}\rangle$.\label{fig:alpha_size}] {\includegraphics[width=0.24\linewidth]{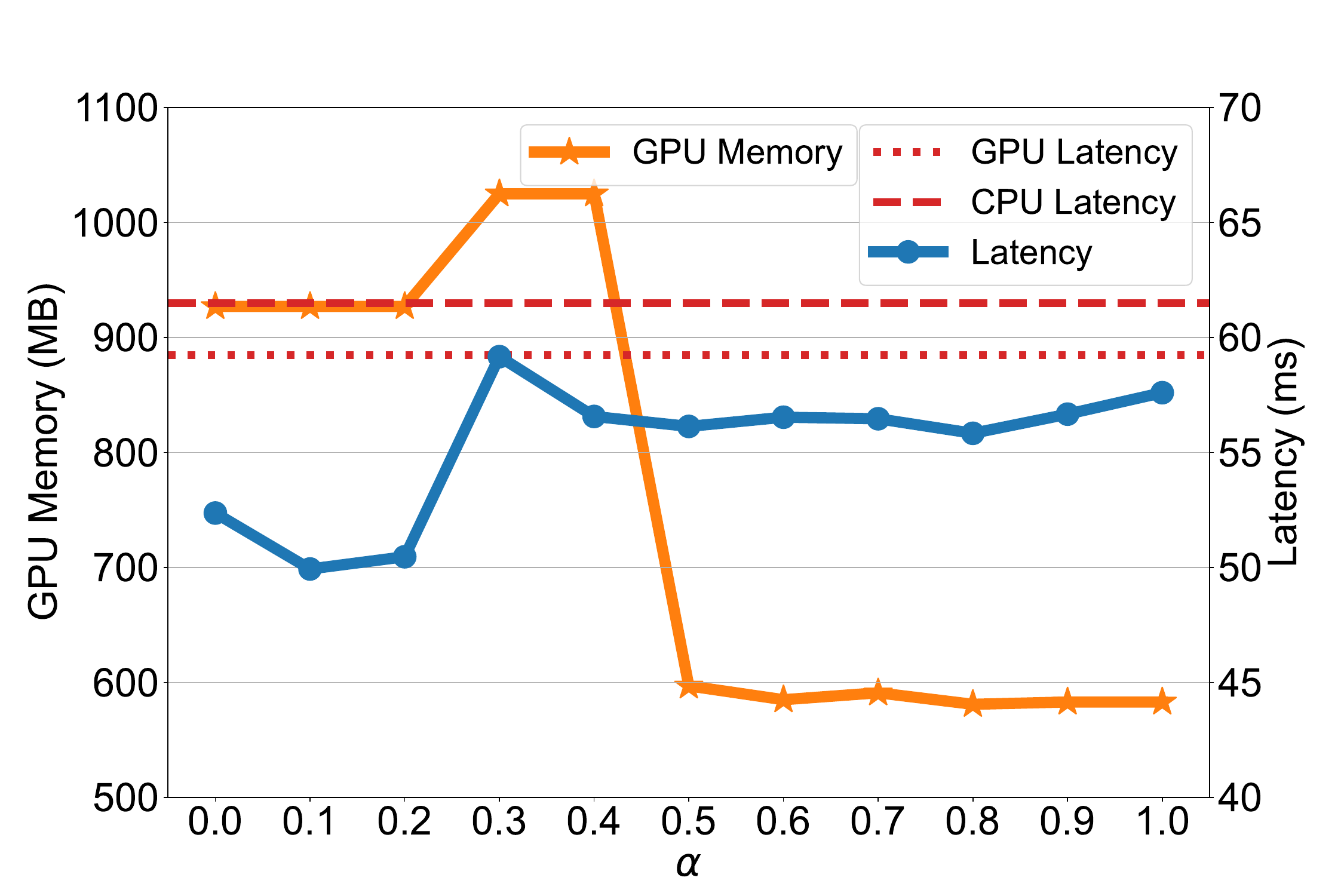}}\hfill
    \subfloat[$\langle{12, 8, 64, 64}\rangle$.\label{fig:alpha_sl}]{\includegraphics[width=0.24\linewidth]{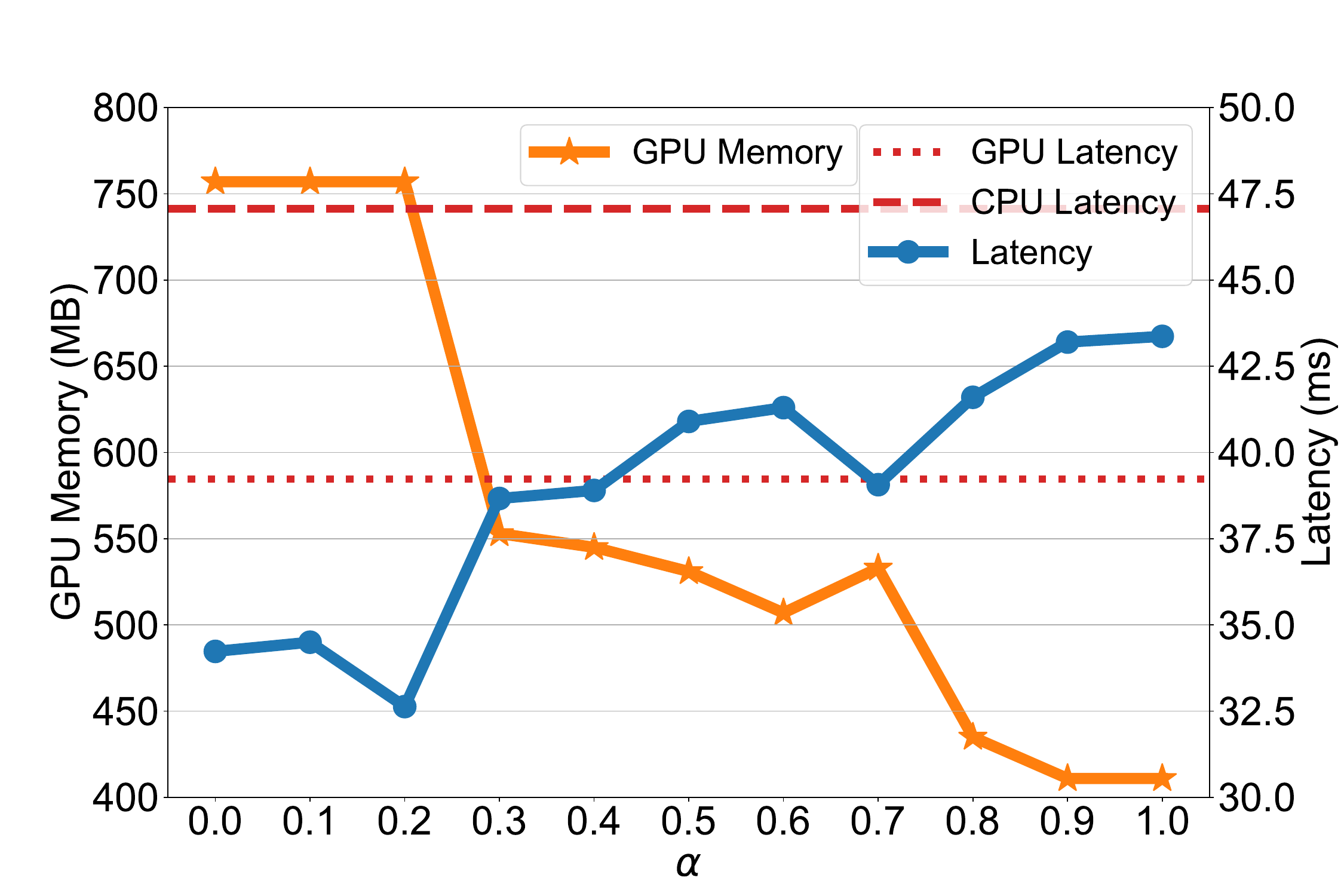}}\hfill
    \caption{$\alpha$'s Effect on Balancing Execution Latency and Memory Requirement. The sub-caption of each sub-figure is a quadruple of $\langle{\mathtt{num\_of\_layers}, \mathtt{batch\_size}, \mathtt{io\_size}, \mathtt{seq\_len}}\rangle$. The orange and blue lines describe the memory and latency as the hyper-parameter $\alpha$ varies, and two red horizontal lines represent the latency of vanilla execution patterns.}
\label{fig:alpha}
\end{figure*}

We first experiment how \name performs under different execution patterns. 
The structure of the baseline model is $\mathtt{num\_of\_layers}=12, \mathtt{batch\_size}=8, \mathtt{io\_size}=64, \mathtt{seq\_len}=96$.
We experiment with four patterns: GPU, CPU, Latency-Optimal and Memory-Optimal, and compare various metrics of different patterns in \tab{e2e}. 
We set $\alpha$ to zero in the Latency-Optimal pattern, and then tune it larger to minimize GPU memory requirement with the guarantee of inference latency in the Memory-Optimal pattern.
We also calculate the reduction rate of GPU memory footprint and execution latency for Latency-Optimal and Memory-Optimal patterns.
Results show that in the Latency-Optimal pattern, \name can not only reduce inference latency by $12.3\%$ from $59.3$ milliseconds to $52.0$ milliseconds, but also decrease GPU memory footprint from $1643.0$ MB to $948.1$ MB by $42.3\%$.
As we relax the latency restrictions without exceeding GPU execution pattern, we can achieve a maximum memory reduction by $61.7\%$.

We further explore whether \name can adapt to different variants of RNNs.
We introduce a wider range of configurations by varying the structural parameters, including number of layers, batch size, input/output dimensions, and sequence length. 
\tab{e2e} shows twelve of such variants where the parameters of these variants are consistent with the basic structure except for the specified ones.
It should be noted that if the latency reduction rate of Latency-Optimal pattern is minus, it means hybrid execution on CPUs and GPUs is not suitable for such an model structure, let alone Memory-Optimal pattern.
Since the number of layers and sequence length play a critical role in model structures, it decides how well the LSTM can collaboratively utilize CPUs and GPUs for inter-operator parallelism.
\tab{e2e} shows that although a smaller number of layers can still achieve memory optimization, it leads to bad performance in latency, no matter in Latency-Optimal or Memory-Optimal patterns, \eg when the number of layers is four.
It is also applicable to explain why a larger sequence length brings little benefit to optimize execution latency.
For the batch size and input/output dimensions, their values determine the computation efficiency of each cell in an RNN on the CPU and GPU.
Since GPUs can take advantage of a larger batch size or input/output dimension to explore its parallelism capacity, \name is not suitable when the batch size is equal to 16 or the input/output dimension is 128 in \tab{e2e}.

%% file: contents/5/5_3_component.tex
\subsection{Effectiveness of $\alpha$\label{sec:alpha}}

We evaluate the effectiveness of $\alpha$ in trading off between inference latency and GPU memory requirement. We experiment with the LSTM model with the baseline structure and three variants, and \fig{alpha} shows the results.
In addition to the GPU memory and execution latency that vary as we increase $\alpha$ from $0.0$ to $1.0$ with the stepsize of $0.1$, we also add execution latency of vanilla execution patterns in the figure. 
\fig{alpha_baseline} is the result with baseline structures, and \fig{alpha_bs}, \fig{alpha_size} and \fig{alpha_sl} are selected from \tab{e2e} with a different batch size, input/output dimension, and dequence length.
As expected, the selection of hyper-parameter determines the preference for latency or memory optimization.
All the sub-figures show a similar tendency, \ie with the increasing of $\alpha$, GPU requirement decreases but the latency keeps increasing.
According to \eq{target}, a larger $\alpha$ denotes we spare more kernels to CPU to explore inter-operator parallelism on multi-core CPUs, but may result in SLO violation becuase of memory movement or long CPU execution. 
When $\alpha$ is equal to $0.3$ for the baseline structure, \name reaches its critical point of memory optimization without violation of SLO.
However, when $\alpha$ is larger than $0.3$, inference latency will exceed GPU, which means \name can not work for further optimization.



%% file: contents/5/5_4_cluster.tex
\begin{figure*}[!t]
    \centering
    \subfloat[$\langle{12, 8, 64, 96}\rangle$.\label{fig:cluster_baseline}]{\includegraphics[width=0.33\linewidth]{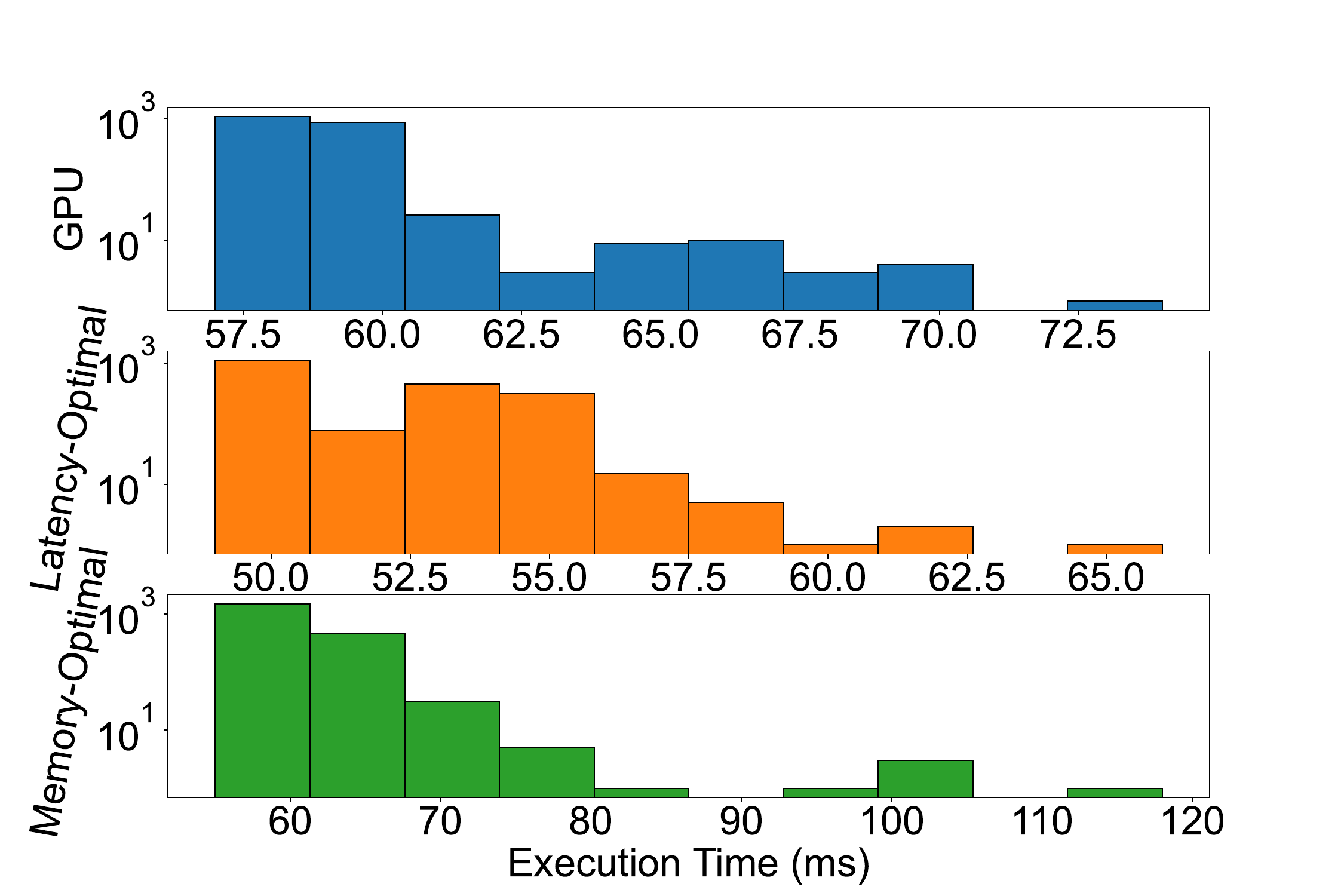}}\hfill
    \subfloat[$\langle{12, 2, 64, 96}\rangle$.\label{fig:cluster_bs=2}]{\includegraphics[width=0.33\linewidth]{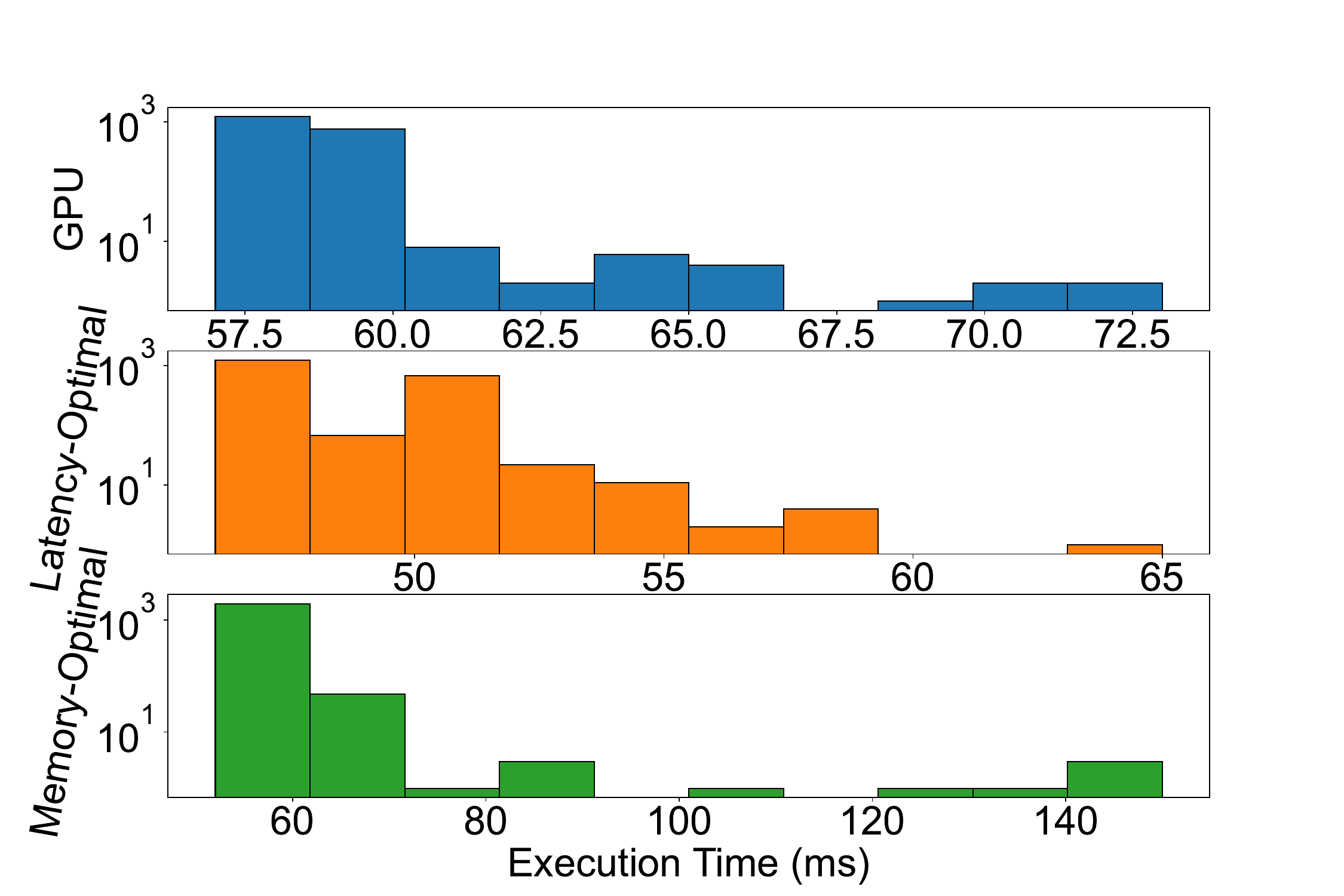}}\hfill
    \subfloat[$\langle{12, 8, 64, 64}\rangle$.\label{fig:cluster_size=64}]{\includegraphics[width=0.33\linewidth]{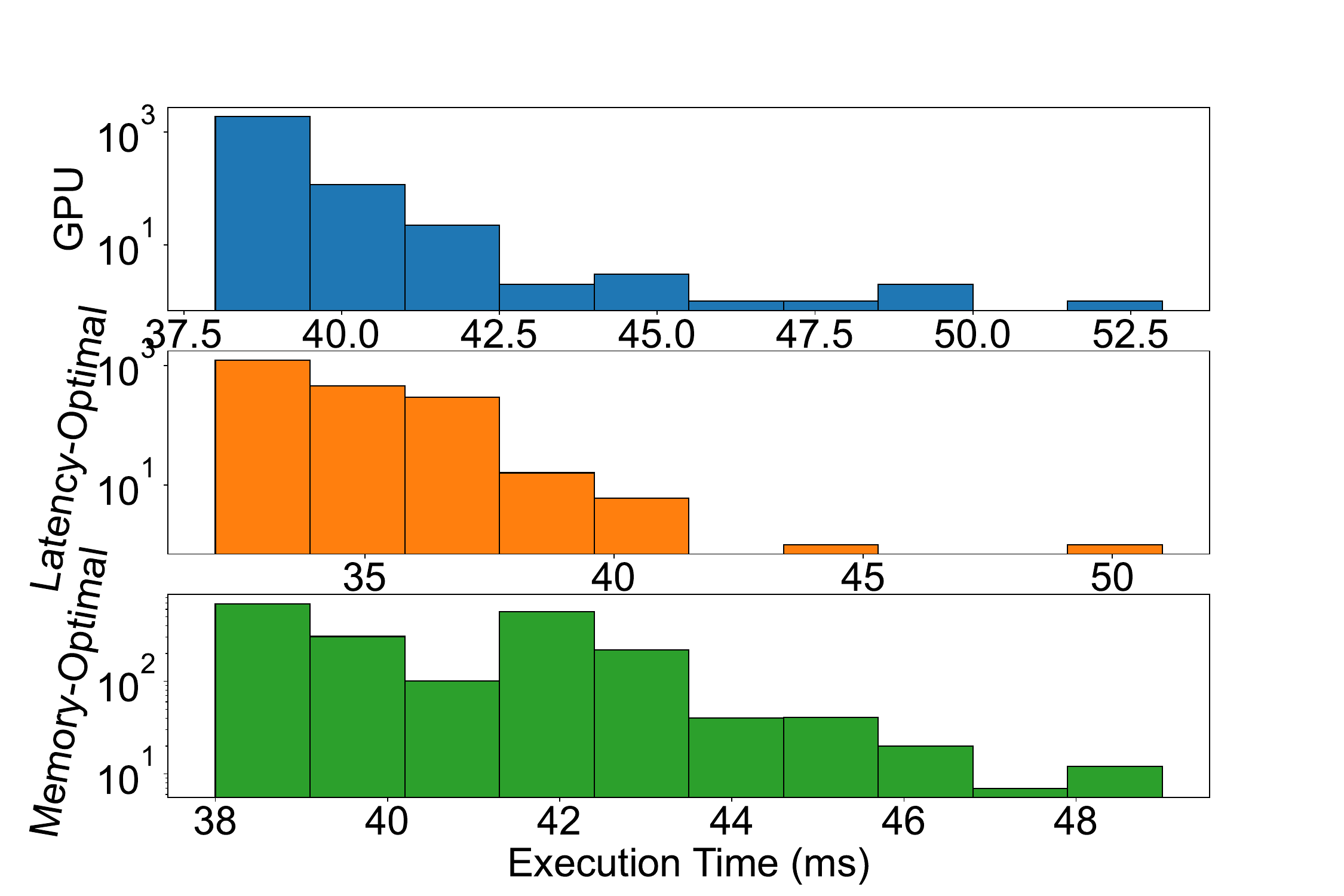}}\hfill
    \caption{Histogram of Execution Time in Cluster Evaluation. The sub-caption of each sub-figure is a quadruple of $\langle{\mathtt{num\_of\_layers}, \mathtt{batch\_size}, \mathtt{io\_size}, \mathtt{seq\_len}}\rangle$. Note that the x-axis range of each sub-graph is different, and the y-axis is processed with logarithm for better visual effect.}
\label{fig:cluster}
\end{figure*}

\subsection{Local Cluster Evaluation\label{sec:local}}

To evaluate {\name}'s performance with read-world workloads, we experiment it with local cluster evaluation.
To display {\name}'s impact on SLO, we first select three models including the baseline one from \tab{e2e}.
For each model, we duplicate it for 20 times and serve them in three patterns, GPU, Latency-Optimal and Memory-Optimal.
\fig{cluster} illustrates the histogram of the execution time for each model in different patterns.
Taking our baseline model for example, we find out that the mode of execution time in GPU pattern appears between 57.5 and 60.0, while that in Latency-optimal pattern appears around 50.0.
Since the Memory-Optimal pattern relaxes the execution time restriction, its mode is a little bigger, bringing reduction in memory footprint.
However, we also observe that the Memory-Optimal pattern may bring long tail latency, especially for \fig{cluster_baseline} and \fig{cluster_bs=2}.

\begin{equation}\label{eq:define1}
\mathtt{slo\_violation} = \frac{\mathtt{\#~of~violation}}{{\mathtt{\#~of~invocations}}}
\end{equation}

\begin{equation}\label{eq:define2}
\mathtt{swapping\_rate} = \frac{\mathtt{\#~of~swapping}}{\mathtt{\#~of~invocations}} 
\end{equation}

\begin{table}[!t]
    \centering
    \caption{Local Cluster Evaluation.}
    \label{tab:final-results}
    \resizebox{\linewidth}{!}{%
    \begin{tabular}{@{}ccc@{}}
        \toprule
         & $\mathtt{slo\_violation}$ & $\mathtt{swapping\_rate}$ \\ \midrule
        \textbf{GPU} & 3.4\% & 44\% \\
        \textbf{Latency-Optimal} & 1.2\% & 9\% \\
        \textbf{Memory-Optimal} & 4.2\% & 0 \\ \bottomrule
    \end{tabular}
    }
\end{table}

We further use the nine feasible models listed in \tab{e2e}, and simulate such a workload where each model's requests come uniformly.
Our evaluation metrics include SLO violation rate and model swapping rate defined in \eq{define1} amd \eq{define2}, and \tab{final-results} shows the results of the local custer evaluation.
We first compare the SLO violation rate between GPU, Latency-Optimal and Memory-Optimal patterns. The Latency-Optimal pattern can not only decrease memory footprint but also reduce execution time. But the Memory-Optimal pattern increases $\mathtt{slo\_violation}$ rate becuase of relaxed SLO restriction.
We also compare the model swapping rate when serving multiple deep learning models in one inference server.
Our experiments confirm that it is common of model swapping between CPUs and GPUs because of the limited GPU memory capacity.
Compared with executing the full model on GPU, the Latency-Optimal pattern can reduce the model swapping rate from 44\% to 9\%. If we further relax the latency restriction with SLO guarantee, the model swapping rate can decrease 0 because the server can keep all the nine models in GPU memory.

%% file: contents/06_related.tex
\section{Related Work}

\paragraph{Heterogeneous Computing.} Modern computing clusters are equipped with heterogeneous platforms, \eg CPUs and GPUs for deep learning workloads.
Many existing works focus on resource management and scheduling to explore their computation potential.
CHARM~\cite{zhang2021charm} provides collaborative resource management between CPUs and GPUs for latency-critical tasks.
It achieves a trade-off between resource consumption and SLO satisfaction by dynamically adjusting resource quota among platforms.
Allox~\cite{le2020allox} is designed based on the performance gap of heterogeneous platforms for different model structures.
Dopia~\cite{cho2022dopia} aims to improve the performance of data-intensive workloads by resolving the limited memory bandwidth on integrated architectures.

\paragraph{LSTM Application and Optimization.} \textbf{}LSTMs have extensive applications in various fields due to their capability of capturing the correlation between time-series data.
In finance, since the financial market is a process of time series change, LSTMs can be used to capture the characteristics of the market and make predictions.
For example, LSTMs are utilized for prediction of stock market prices~\cite{bukhari2020fractional, liu2018stock} and financial fraud detection~\cite{alghofaili2020financial} to improve business security.
In biology, LSTMs are integrated with CNNs to predict protein structures~\cite{cheng2020protein}, diabete detection~\cite{rahman2020deep}, inter-protein interaction~\cite{ahmed2019identifying}.
Besides, LSTMs are also applied in transportation, for example, to predict the traffic flow during a continuous period of time~\cite{zhao2017lstm, bi2021hybrid}.

Because of its widespread application, optimizing its computation efficiency on heterogeneous platforms is necessary.
GRNN~\cite{holmes2019grnn} is an RNN inference library to improve data reuse and alleviate synchronization overhead when serving RNNs on GPUs.
Zhang \etal identify poor data reuse as the root cause of high execution latency, and design DeepCPU~\cite{zhang2018deepcpu}, a CPU-based RNN serving library claimed to speed up RNN inference by ten times.
Since batching is a common technique to improve system efficiency, BatchMaker~\cite{gao2018low} proposes the technique of cellular batching to improve RNN inference throughput. 
\name reduces GPU memory footprint when serving RNNs and empowers CPU collaboratively for forward computation, which is orthogonal to these related works.

%% file: contents/07_conclusion.tex
\section{Conclusion}

Developing high-performance and cost-efficiency deep learning serving systems is critical to machine learning practitioners.
However, existing serving systems face severe challenges, including GPU memory bottleneck and under-utilization of CPUs.
We observe the opportunity of providing computation power using idle CPUs for inter-operator parallelism across heterogeneous platforms when handling inference workloads,
and design \name, an optimized serving system for RNNs by collaboratively utilizing CPUs and GPUs for forward computation.
By scheduling partial operators to CPUs, \name reduces the GPU memory footprint to serve RNNs within the guarantee of SLO and improves CPU utilization.
Our end-to-end experiments show that \name can reduce inference latency by at most 19.4\% and GPU memory requirement by at most 67.5\% for multi-layer RNNs.